\documentclass{JHEP3}
\usepackage{amsmath}
\input epsf
\usepackage{epsfig}
\usepackage{amssymb}
\usepackage{graphics}
\usepackage[active]{srcltx}

\newcommand{\bea}{\begin{eqnarray}}
\newcommand{\eea}{\end{eqnarray}}
\newcommand{\bean}{\begin{eqnarray*}}
\newcommand{\eean}{\end{eqnarray*}}
\newcommand{\nn}{\nonumber \\}

\def\O #1{\overline{#1}}

\def\W #1{\widetilde{#1}}
\def\WH #1{\widehat{#1}}
\def\und #1{\underline{#1}}

\def\braket#1{\left\langle #1 \right\rangle}

\def\ket#1{\left| #1\right\rangle}
\def\gb #1{ \left\langle #1 \right]}
\def\tgb #1{ \left[ #1 \right\rangle}

\def\Tr{\mathop{\rm Tr}}

\def\eref#1{(\ref{#1})}

\def\a{{\alpha}}

\def\b{{\beta}}

\def\la{\lambda}
\def\eps{\epsilon}

\def\vev{\braket}
\def\tgb #1{ \left[ #1 \right\rangle}
\def\bket#1{\left| #1\right]}
\def\bvev#1{\left[ #1 \right]}
\def\Spaa{\vev}
\def\Spbb{\bvev}
\def\Spab{\gb}
\def\Spba{\tgb}

\def\Label#1{\label{#1}%
  \smash{\hbox to0pt{\raise1ex\hbox{\tiny[#1]}\hss}}}

\title{Recursion Relation for Boundary Contribution }
\author{ Qingjun Jin $^{a}$, Bo Feng$^{ab}$, \footnote{The
unusual ordering of authors is just to let authors get proper
recognition of contributions under outdated practice in China. }\footnote{emails: qingjun@zju.edu.cn; b.feng@cms.zju.edu.cn}
 \bigskip\\
{$^a$\small Zhejiang Institute of Modern Physics, Zhejiang
University, Hangzhou, 310027, P. R. China\\$^b$\small Center of
Mathematical Science, Zhejiang University, Hangzhou, China \\
 }}
\date{\today}
\abstract{It is well known that under a BCFW-deformation, there is a boundary contribution
when the amplitude scales as $\mathcal{O}(z^0)$ or worse. We show that
boundary contributions have a similar recursion relation as scattering amplitude. 
Just like the  BCFW
recursion relation, where scattering amplitudes are expressed as the products of two on-shell sub-amplitudes (plus  possible
boundary contributions),  our new recursion relation expresses
boundary contributions  as products of sub-amplitudes
and boundary contributions with less legs, plus yet another
possible boundary contribution. In other words, the complete scattering amplitude, including boundary contributions, can be obtained by multiple steps of recursions, unless the boundary contributions are still non-zero when all possible deformations are exploited. We demonstrate this algorithm by several examples. Especially, we show
that for standard model like renormalizable theory in 4D, i.e., the theory including only gauge boson, fermions and scalars, the complete amplitude can always be computed by at most four recursive steps using our algorithm.

}


\begin{document}

%%%%%%%%%%%%%%%%%%%%%
\section{The introduction}
%%%%%%%%%%%%%%%%%%%%%%%

In recent years BCFW recursion relation
\cite{Britto:2004ap,Britto:2005fq} has become a standard method to
compute tree-level scattering amplitudes. In its original form, BCFW
recursion relation was presented for 4d Yang-Mills theory in the language of
spinors, but soon the method were applied to various other theories\footnote{
For more information,
see reviews \cite{Bern:2007dw, Feng:2011np, Elvang:2013cua} and references therein.}. Despite its successes, BCFW recursion relation has met  difficulties applying to certain theories\footnote{A typical example is  when
all external particles are scalars and fermions in the Standard
Model. } whose amplitudes do not have the desired vanishing scaling
in the large limit of deformation parameter. A naive application of
BCFW recursion relation fails to capture a piece of amplitude
(usually called boundary contributions), which corresponds to the
residue at infinity.
%In \cite{Cheung:2008dn} Cheung proved an 
%amplitude always satisfies recursion relations if it contains a
%gluon leg, but a Standard Model amplitude may not contain any gluon legs.

Several proposals have been made to find boundary contributions. The
first \cite{Paolo:2007,Boels:2010mj} is to introduce auxiliary
fields so that in the enlarged theory, there are no boundary
contributions.  The second
\cite{Feng:2009ei,Feng:2010ku,Feng:2011twa} is to carefully analyze
Feynman diagrams and then isolate their boundary contributions,
which can be evaluated directly or recursively afterwards.  The
third \cite{Benincasa:2011kn, Benincasa:2011pg, Feng:2011jxa} is to
express boundary contributions in terms of roots of amplitudes.
These three methods are, however, effective only for limited types of theories. Recently a systematical algorithm, based on carefully analysis of pole structure of boundary contributions, has been proposed in \cite{Feng:2014pia}. Though in principle the method is applicable to any quantum field theory, in practice it suffers from high computational complexity.

In this paper, we present a new method to
compute boundary contributions. The key observation is that with
properly chosen deformations, boundary contributions satisfy similar
recursion relations as scattering amplitudes. Just like the  BCFW
recursion relation, where scattering amplitudes are expressed as (a
sum of) the products of two on-shell sub-amplitudes (plus  possible
boundary contribution),  our new recursion relation expresses
boundary contributions  as (a sum of) products of sub-amplitudes
and boundary contributions with less legs, plus yet another
possible boundary contribution. The new boundary contribution is
subsequently computed by a new shift, and the recursion ends
whenever the remaining boundary contribution vanishes. This multi-step
recursion is (almost) as efficient as BCFW recursion, but applicable
to more general models.

The paper is organized as follows. In section 2 after a short
discussion of pole structure, we present our main result: the recursion relations
for boundary contributions.
 In section 3 we show that a pure scalar  $\phi^m$ theory amplitude can be
computed via a $(m-1)$ step recursion. In section 4 we analyze all
possible  boundary contributions of Standard Model like theories,
and show that any amplitude in this theory can be computed via a (at
most) 4 step recursion. In section 5, we present two explicit
examples using our method. In Appendix A, we discuss some
mathematical aspects of multi-variable integrations. In Appendix B,
we present recursion relation for boundary contributions under other
choices of deformations. In Appendix C, propagator in light-cone
gauge has been discussed.

%%%%%%%%%%%%%%%%%%%%
\section{The recursion relation for boundary contribution}
%%%%%%%%%%%%%%%%%%%

The key idea of BCFW recursion is determining scattering amplitudes by their poles. In order to find a recursion relation of boundary contributions, we also need to be very clear about the poles of the boundary contribution. First consider the primary deformation (BCFW-deformation) $\Spab{1|n}$,
\bea \la_1\to \la_1-z\la_n,~~~~\W\la_n\to \W\la_n+z\W\la_1~.~~\label{BCFW-1n}\eea
Let us use indices $I,J$ to denote  subsets of remaining particles ${\cal T}\equiv\{2,3,...,n-1\}$. 
For later convenience, we also define $q_i^{\mu}=\frac{1}{2}[i|\gamma^{\mu}|n\rangle$, then \eqref{BCFW-1n} can be written as
\begin{equation}
p_1\rightarrow p_1-zq_1,\ p_n\rightarrow p_n+zq_1\ .
\end{equation}
Under the deformation,
the expression of tree-level amplitudes coming from Feynman diagrams will be
\bea A(z) & = & { f(z)\over \prod_{I\subset {\cal T}} P_I^2 \prod_{J\subset {\cal T}} (P_J+p_1-zq_1)^2 }\nn
& = & \sum_{J\subset {\cal T}} {R_{J}\over (P_J+p_1-zq_1)^2 }+ B^{\langle1|n]}+C_1 z+C_2 z^2+\cdots~~~\label{Az-split}\eea
where $R_J$'s are the residues  of corresponding poles and $B^{\langle 1|n]}$ is the boundary contribution we want to find.

 To read out $B^{\langle 1|n]}$, a good way is to do the large $z$ expansion in the first line of \eref{Az-split}.
Using
\bea {1\over (P_J+p_1-zq_1)^2 } & = &{1\over -z\Spab{n|P_J+p_1|1}}\sum_{i=0}^{\infty}  \left({(P_J+p_1)^2\over z\Spab{n|P_J+p_1|1}}\right)^i
~~~~\label{BD-exp-0}\eea
we found that
\bea A(z)
& = & { f(z)\over \prod_{I\subset {\cal T}} P_I^2} \prod_{J\subset {\cal T}}\left[ {1\over -z\Spab{n|P_J+p_1|1}}\sum_{i=0}^{\infty}  \left({(P_J+p_1)^2\over z\Spab{n|P_J+p_1|1}}\right)^i\right]~~\label{BD-exp}\eea
and $B^{\langle 1|n]}$ can be read out by selecting same power of $z$ in numerator $f(z)$ and denominators. In other words,   poles of $B$ can be
\bea  P_{I\subset {\cal T}}^2, ~~~~~~~\Spab{n|P_{J\subset {\cal T}}|1}^a~~~~\label{BPole-1}\eea
It is worth to notice that, in principle, $B^{\langle 1|n]}$ can have terms which are  pure {\sl polynomials in momentum} (i.e. they do not have any pole), and our method is not applicable. This can happen in many effective theories with higher dimension operators.

%%%%%%%%%%%%%%%%%%%%%%%%%
\subsection{Recursion relation for boundary contribution}
%%%%%%%%%%%%%%%%%%%%%%%%

As discussed  in \cite{Feng:2014pia}, we can use a different deformation  to compute $B^{\langle 1|n]}$. Without loss of generality, we will choose the deformation $\langle 2|n]$. A crucial merit of this deformation is that spurious poles $\Spab{n|P_{J\subset {\cal T}}|1}$ in \eref{BPole-1} as well as others $\Spab{n|P_{J\subset {\cal T}}|i}$ generated in middle steps are invariant under the deformation. In other words, under this deformation  only physical single poles $P_{I\subset {\cal T}}^2$ in \eref{BPole-1} are detected.

Following the proof of BCFW recursion relations, we evaluate  the contour integration
\bea B^{\langle 12|n]}=
\frac{1}{2\pi i}\oint_{|w|=R} dw {B^{\langle 1|n]}(w)\over w}
=B^{\langle 1|n]}+\sum_{w^*}{\rm Res} \left( {B^{\langle 1|n]}(w)\over w}\right)_{w=w^*}
~.~~~\label{Bound-rec-1}\eea
where $B^{\langle 12|n]}$ is the possible remaining boundary contribution and the residue part is given by recursion relation
\bea \boxed{-{\rm Res} \left( {B^{\langle 1|n]}\over w}\right)_{w=w_{\cal I}} =  \sum_h A_L(\WH p_2(w_{\cal I}),{\cal I},-P^h(w_{\cal I})) {1\over (p_2+P_{\cal I})^2} B^{\langle 1|n]}( p_1, \WH p_n(w_{\cal I}), \O {\cal I},P^{-h}(w_{\cal I}))} ~~~~\label{Bound-rec-3} \eea
with $w_{\cal I}= {(p_2+P_{\cal I})^2\over \Spab{n|P_{\cal I}|2}}$  and\footnote{Here $1\notin {\cal I}\bigcup \O{\cal I}$, because according to \eqref{BPole-1}, $B^{\langle 1|n]}$ does not have poles at $(p_1+p_2+P_J)^2$.} ${\cal I}\bigcup \O{\cal I}=\{3,4,...,n-1\}$. In \eref{Bound-rec-3}, the $B^{[1|n\rangle}( p_1, \WH p_n(w_{\cal I}), \O {\cal I},P^{-h}(w_{\cal I}))$ is the boundary contribution of lower point amplitudes under deformation $\Spab{1|n}$. 
%It is worth to mention that naively under the deformation $\Spab{2|n}$, we should detect pole $\Spaa{1|2-z n}$ and $(p_2(z)+p_1+P_J)^2$. However, poles $\Spaa{1|2}$ and $(p_2+p_1+P_J)^2$  have been detected under the $\Spab{1|n}$-deformation, thus by our discussion in \eref{BD-exp} and \eref{BPole-1}, $B^{[1|n\rangle}$ does not depend on poles $\Spaa{1|2}$ and $(p_2+p_1+P_J)^2$. In other words,  we should not consider the case $1 \in {\cal I}$ in \eref{Bound-rec-3}.

\begin{figure}[htb]
\centering
\includegraphics[scale=0.8]{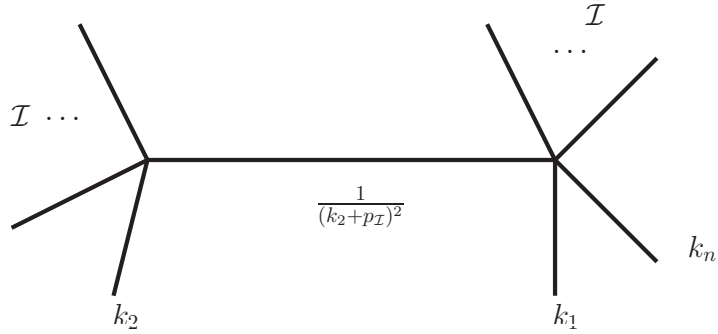}
\caption{Recursion relation of $B^{\langle1|n]}$ using the $\langle 2|n]$ deformation. The left hand side of the diagram is an on-shell sub-amplitude, while the right hand side of the diagram is the boundary $B^{\langle1|n]}$ with less legs.}
\label{fig:boundary1n}
\end{figure}

To prove \eref{Bound-rec-3}, first we notice that
\bea B^{\langle 1|n]}(p_1,p_2,...,p_{n-1},p_n)=\frac{1}{2\pi i}\oint_{|z|=R} {dz\over z} A(p_1-zq_1,p_2,..., p_{n-1},
p_n+zq_1)~~~\label{B01-rep}\eea
thus
\bea & & -{\rm Res} \left( {B^{\langle 1|n]}\over w}\right)_{w=w_{\cal I}}  = - \frac{1}{2\pi i}\oint_{w=w_{\cal I}} {dw\over w}
B^{\langle 1|n]}(p_1,p_2-wq_2,..., p_{n-1},p_n+wq_2)\nn
& = & - \frac{1}{(2\pi i)^2}\oint_{w=w_{\cal I}} {dw\over w} \oint_{|z|=R} {dz\over z} A(p_1-zq_1,p_2-wq_2,p_3,..., p_{n-1},
p_1+zq_1+wq_2)~~~\label{B01-rp-1}\eea
where we have used \eref{B01-rep} at the second line.
Above integration can be parameterized by $w= w_{\cal I}+\eps e^{i\a}$ and $z= R e^{i\b}$, thus the contour integration becomes following double integrations
\begin{equation}\label{B01-rp-2}
 - \frac{\epsilon}{(2\pi )^2}\int_0^{2\pi}  \frac{d\alpha d\beta }{w_{\cal I}+\eps e^{i\a}}
A(p_1-R e^{i\b}q_1,p_2-(w_{\cal I}+\eps e^{i\a})q_2,p_3,..., p_{n-1},
p_1+R e^{i\b}q_1+(w_{\cal I}+\eps e^{i\a})q_2)
\end{equation}

Now for $R$ big enough but finite and $\eps$ small enough but finite, $A$ is finite (i.e., there is no pole along the integral path). Using the {\bf Fubini-Tonelli theorem} reviewed in Appendix \ref{appendixA}, we can exchange the ordering of two integrations, thus \eref{B01-rp-1} becomes
\bea & &  - \frac{1}{(2\pi i)^2} \oint_{|z|=R} {dz\over z}\oint_{w=w_{\cal I}} {dw\over w} A(p_1-zq_1,p_2-wq_2,p_3,..., p_{n-1},
p_1+zq_1+wq_2)\nn
& = & \frac{1}{2\pi i}\sum_h A_L(\WH p_2(w_{\cal I}),{\cal I},-P^h(w_{\cal I})) {1\over (p_2+P_{\cal I})^2} \oint_{|z|=R} {dz\over z}A_R (p_1-zq_1, \O {\cal I},p_n+zq_1+w_{\cal I}q_2, P^{-h}(w_{\cal I} )) \nn
& = & \sum_h A_L(\WH p_2(w_{\cal I}),{\cal I},-P^h(w_{\cal I})) {1\over (p_2+P_{\cal I})^2} B^{\langle1|n]}( p_1, \WH p_n(w_{\cal I}), \O {\cal I},P^{-h}(w_{\cal I}))~~~\label{B01-rp-2}\eea

Thus we have proved \eref{Bound-rec-3}.
If $B^{\langle 12|n]}\neq 0$, we can take a third deformation, for example $\Spab{3|n}$. First we write
\bea B^{\langle12|n]} & = &\frac{1}{(2\pi i)^2} \oint_{|z_2|=R_2} {d z_2\over z_2}\oint_{|z_1|=R_1} {d z_1\over z_1} 
A(p_1-z_1q_1, p_2-z_2q_2,\cdots,p_n+z_1q_1+z_2q_2)~~~\label{B12-rep}\eea
Using the contour integration, we obtain
\bea B^{\langle12|n]} & = & B^{\langle123|n]}-\sum_{w_{\cal I}}
 {\rm Res} \left( {B^{\langle12|n]}(w) \over w}\right)_{w_{\cal I}}  \eea
It is important to emphasize that since above two integrations are around infinity, in general we can not change the ordering (see the discussion in the Appendix \ref{appendixA}), i.e., $B^{\langle12|n]}\neq B^{\langle21|n]}$.  Nevertheless, we can still change the order of $w$ and $z_i$ integration, and we find
\bea & & -
 {\rm Res} \left( {B^{\langle12|n]} \over w}\right)_{w_{\cal I}}= A_L(p_3-w_{\cal I}q_3,
{\cal I}, -P^{h}) {1\over (p_3+P_{\cal I})^2} B^{\langle12|n]}(P^{-h}, \O {\cal I}, p_n+w_{\cal I}q_3),~~~~1,2\not\in {\cal I}~~~~~~~\label{B123-2}\eea

Before ending this section, let us give some remarks. For the application of above result, it seems crucial that there is a choice such that after finite steps, we should have $B^{\langle 1\cdots k|n]}=0$. In later part of the paper, we will discuss several theories in which such a choice always exists. But there are theories in which the boundary contributions do not vanish after exploiting all shifts. If we define the recursion part of the $i$-th deformation as $A^{\langle 1\cdots i|n]}$, then we have
\begin{equation}\label{amexpansion}
A=A^{\langle 1|n]}+A^{\langle 12|n]}+\cdots+A^{\langle 1\cdots n-2|n]}+B^{\langle 1\cdots n-2|n]}\ .
\end{equation}
In order for our algorithm to be complete, an efficient method to determine the last boundary contribution, $B^{\langle 1\cdots n-2|n]}$, is desirable. At the same time, it is equally important to explore whether the later terms in \eqref{amexpansion} are suppressed(for example by some large energy scale). In this case one can use the first several terms as a good approximation of the complete amplitude.

%%%%%%%%%%%%%%%%%%%%%
\section{Scalar Theory}
%%%%%%%%%%%%%%%%%%%%%

Starting from this section, we will demonstrate our algorithm by several examples. The first simple example is the  real scalar theory with $\phi^m$ interaction term, i.e.,
the Lagrangian is given by
\begin{equation}
L=-\frac{1}{2}\partial_{\mu}\phi^I\partial^{\mu}\phi^I+\frac{\sigma}{m!}\phi^m~.~~\label{Phi-Lag}
\end{equation}
The vertex $\phi^m$ will contribute possible boundary terms for $n$-point amplitude when $n\geq m$.
For $n=m$, the contribution is just $\sigma$ and we could not detect it using pole. Thus we will
consider the case $n>m$.  In fact, to get nontrivial Feynman diagrams, we need to have $n=2+(m-2)V$ where
$V$ is the number of vertices.

\begin{figure}[htb]
\centering
\includegraphics[scale=1]{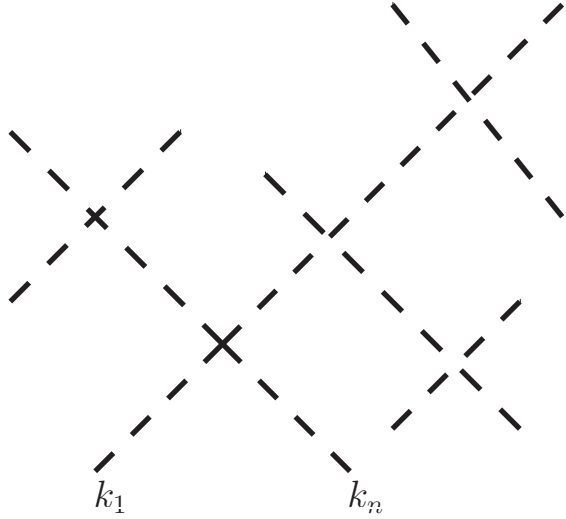}
\caption{An example of $\mathcal{O}(z^0)$ diagrams under $\langle 1|n]$ deformation in $\phi^4$ theory.}
\label{fig:scalarm}
\end{figure}

It is easy to see that under the primary deformation $\Spab{1|n}$, the boundary contribution comes from Feynman diagrams where $1,n$  attach to same vertex (see Figure \ref{fig:scalarm}).  If
we define a non-overlapping $(m-2)$-splitting $K$ of the set $\{2,3,\cdots,n-1\}$ as
\begin{equation}
K=\{K_1,K_2\cdots K_k\}
\end{equation}
with each $K_i$ having at least one element (the ordering does not matter in the splitting), the boundary
contribution is given by
 \begin{equation}
B^{\langle 1|n]}=\sigma \sum_{K\in \Lambda}\frac{1}{p_{K_1}^2\cdots p_{K_{m-2}}^2}\mathcal{A}_1(K_1, P_1)\cdots \mathcal{A}_{m-2}(K_{m-2}, P_{m-2})~~~~\label{Sm-Boundary}
\end{equation}
where $\Lambda$ is the set of all allowed splitting and  $P_{i=1,...,m-2}$ are, in fact, inner particles in  Figure \ref{fig:scalarm}. From \eref{Sm-Boundary}, it is easy to see that since $n>m$, there are at least
two vertices. Thus for worst diagrams, where  there is only one inner particle connecting to the vertex attached by $1,n$, at most $(m-2)$ new deformations of $\Spab{i|n}$ type besides $\Spab{1|n}$ will be enough to completely determine the boundary contribution $B^{\langle 1|n]}$.
For example,  for $\phi^4$ theory, under the $\Spab{1|n}$ ($n>4$), boundary part are given by Feynman diagrams
where $1,n$ attach to same vertex. Under the second deformation $\Spab{2|n}$, only these Feynman diagrams
where $1,2,n$ attach to same vertex are undetected, but they will be detected by the third deformation, for example, $\Spab{3|n}$. Thus by total three steps we can determine the full amplitude.

It is worth to mention that our above discussion of pure scalar theory does not depend on the detail if the theory contains  lower point vertex $\phi^{p}$ with $p<m$ in the Lagrangian.

%%%%%%%%%%%%%%%%%%%%%
\section{The standard model like theory}
%%%%%%%%%%%%%%%%%%%%%%

In this section, we discuss standard model like theory, for which
the Lagrangian is given by\footnote{In principle, one could add
terms ${1\over 3}  a_{IJK} \phi^I\phi^J\phi^K$ into Lagrangian, but
as one can check,  our following discussion will not be modified. }
\begin{equation}
\begin{aligned}
L=&\Tr\left(L_0-\frac{i}{\sqrt{2}}(L_Y-\bar{L}_Y)+L_{\phi}\right),\\
L_0=&-\frac{1}{4}F_{\mu\nu}F^{\mu\nu}-\frac{1}{2}D_{\mu}\phi^I D^{\mu}\phi^I+i\bar{\psi}_A\bar{\sigma}^{\mu}D_{\mu}\psi^A,\\
L_Y=&\lambda_{IAB}\phi^I\psi^A\psi^B
,~~~\ \bar{L}_Y=\bar{\lambda}_I^{BA}\phi^I\bar{\psi}_A\bar{\psi}_B,~~~\bar{\lambda}_I^{AB}=(\lambda_{IAB})^*\\
L_{\phi}=&\frac{1}{4}a_{IJKL}\phi^I\phi^J\phi^K\phi^L.\\
\end{aligned}~~~\label{deforml}
\end{equation}
For simplicity, the gauge group is $SU(N)$ and  scalars and fermions
are massless.  For this theory, we can classify configurations of
external particles as following: (a) there is at least one gluon;
(b) there is no gluon, but at least four fermions; (c) there are
only two fermions and others are scalars; (d) all are scalars.

For the case (a), as has been proved in
\cite{ArkaniHamed:2008yf,Cheung:2008dn}, if external  particles
contain  at least one gluon, there is always a good deformation
without boundary contribution, so we will not consider case (a)
further.

For case (b), there is also a one-step deformation to completely
determine amplitude.  If there is at least
three positive  fermions, for example, $1,2,3$, we can make
following Risager deformation \cite{Risager:2005vk},
\begin{equation}
\begin{aligned}
|1\rangle\rightarrow &|1\rangle+z[23]|\eta\rangle,\\
|2\rangle\rightarrow &|2\rangle+z[31]|\eta\rangle,\\
|3\rangle\rightarrow &|3\rangle+z[12]|\eta\rangle.\\
\end{aligned} ~~~\label{4fermion-Risager}
\end{equation}
\begin{figure}[htb]
\centering
\includegraphics[scale=1]{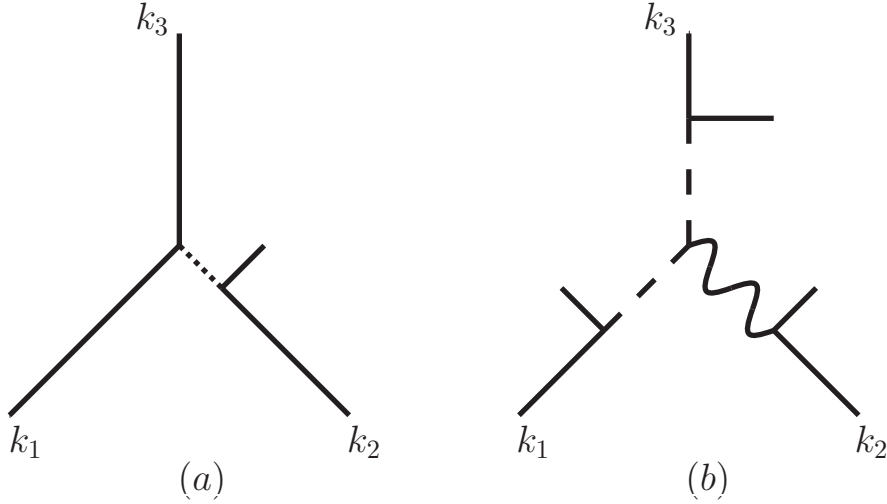}
\caption{Risager deformation. In case (a), the vetex in the center scales as $z^0$, but it is attached to a $\mathcal{O}(z^{-1})$ propagator. In case (b), the vertex in the center scales as $z^1$, but it is attached to three $\mathcal{O}(z^{-1})$ propagators.}
\label{fig:risagershift}
\end{figure}
Now we count the power of $z$. First the wave functions of three
particles are $\bket{1},\bket{2},\bket{3}$, so they scale as $z^0$.
Next, each fermionic propagator scales as $z^0$ while each
bosonic propagator scales as $z^{-1}$. Last, among the vertices, only
$A^3, A\phi\partial \phi$ contribute $z^1$ factors.  But since these
vertices are attached to at least two\footnote{A vertex can be attached to three deformed bosonic propagators, e.g. the vertex in the center of Figure \ref{fig:risagershift}(b).} bosonic propagators, thus the number of
vertices is always less than the number of bosonic propagator (see,
Figure \ref{fig:risagershift}), thus under the large $z$ limit, the integrand vanishes. If
there are two positive fermion $\psi_1,\psi_2$ and two negative
fermion $\psi_3,\psi_4$, we can do following deformation,
\begin{equation}
\begin{aligned}
|1\rangle\rightarrow &|1\rangle+zx_1|3\rangle,\\
|2\rangle\rightarrow &|2\rangle+zx_2|3\rangle,\\
|3]\rightarrow&|3]-z(x_1|1]+x_2|2]).\\
\end{aligned}~~~\label{4fermion-Risager-like}
\end{equation}
Since under this deformation, the wave functions of $\psi_i,i=1,2,3$
are not changed, the power counting of $z$ is similar to the case
where all $\psi_i,i=1,2,3$ are positive. The deformation
\eref{4fermion-Risager-like} is kindly of the union of two
BCFW-deformation using same $z$ variable.

In the next several subsections we will show that for cases (c) and (d), the amplitude can be computed by a (at most) 4 step recursion.

%%%%%%%%%%%%%%%%%%%%%%%
\subsection{The case (c) with only two fermions}
%%%%%%%%%%%%%%%%%%%%%%%%

Now we consider $n$-point amplitude with $2$ fermions and $(n-2)$
scalars. With out loss of generality we assume the particles $1,2$ are scalars while $n$ is the
negative fermion, so its wave function is given by $\ket{n}$. We will work in light cone gauge, which is most convenient for the analysis of boundary behavior.

Since we are mainly interested in gauge coupling, we will neglect Yukawa and quartic scalar coupling terms in \eqref{deforml} for now. The
Light-cone gauge Lagrangian is given by
\begin{equation}
\begin{aligned}
L_{(2)}=&A^{\bar{h}}\partial^2A^h-\frac{1}{2}\partial_{\mu}\phi^I\partial^{\mu}\phi^I
+i\bar{\psi}_A\bar{\sigma}^{\mu}\partial_{\mu}\psi\\
L_{(3)}=&-2ig\left(\frac{\partial^h}{\partial^-}A^{\bar{h}}\right)[A^{\bar{h}},\
\partial^-A^h]
-2ig\left(\frac{\partial^{\bar{h}}}{\partial^-}A^h\right)[A^h,\ \partial^-A^{\bar{h}}]\\
&+gA^h\left(\mathcal{J}^{\bar{h}}-\frac{\partial^{\bar{h}}}{\partial^-}\mathcal{J}^-\right)+gA^{\bar{h}}\left(\mathcal{J}^h
-\frac{\partial^h}{\partial^-}\mathcal{J}^-\right)\\
L_{(4)}=&2g^2\frac{1}{\partial^-}[A^h,\ \partial^-A^{\bar{h}}]\frac{1}{\partial^-}[A^{\bar{h}},\ \partial^-A^h]-\frac{g^2}{2}(\frac{1}{\partial^-}\mathcal{J}^-)^2\\
&+g^2\frac{1}{\partial^-}\mathcal{J}^-
\left(\frac{-i}{\partial^-}[A^h,\ \partial^-A^{\bar{h}}]-\frac{i}{\partial^-}[A^{\bar{h}},\ \partial^-A^h]\right)\\
\mathcal{J}^{\mu}=&ig[\phi^I,\ \partial^{\mu}\phi^I]
-g[\bar{\psi}_A\bar{\sigma}^{\mu},\ \psi^A]\\
\end{aligned}~~~~\label{Light-cone}
\end{equation}
where $A^+=\O q\cdot A$, $A^-=q\cdot A$, $A^h=h\cdot A$ and $A^{\O
h}=\O h\cdot A$ with the basis $q,\O q,h,\O h$ defined as
\begin{equation}
\begin{aligned}
q^{\mu}=&\frac{1}{2}[1|\gamma^{\mu}|n\rangle,\
~~\bar{q}^{\mu}=\frac{[n|\gamma^{\mu}|1\rangle}{2k_1\cdot k_n},~~
~~h^{\mu}=\frac{k_n^{\mu}}{k_1\cdot k_n},\
~~\bar{h}^{\mu}=k_1^{\mu}\\
\end{aligned}~~~~~\label{q-basis}
\end{equation}
All inner products of basis vanish except $q\cdot\bar{q}=-1$,
$h\cdot\bar{h}=1$. The advantage of Light-cone gauge Lagrangian
\eref{Light-cone} is that under the deformation $\Spab{i|n}$, many
$z$-factor coming from vertices will be canceled out. For example,
for  $L_{(4)}$ part, only $\partial^-$ operator appears, but it is
equal to $p^-=q\cdot p$, so under the deformation $\Spab{i|n}$, the
$z$-dependent part will be $p^-(z)\sim q\cdot (z q_i)=0$. In
other words, four point vertex in Light-cone gauge will never
contribute $z$ factor. Similar observation can be made for $L_{(3)}$
part. Now we will have $p^h=p\cdot h$ and $p^{\O h}=p\cdot \O h$.
Under the deformation, we will have $(z q_i)\cdot h=0$
although $(z q_i)\cdot \O h\neq 0$ when $i\neq 1$. In other
words, under the $\Spab{i|n}$-deformation, only $A^h A^{\O h} A^{\O
h}$ vertex, $\phi \phi A^{\O h}$ vertex and $A^{\O h} \psi^+\psi^-$
vertex contribute factor $z$. We will use this important observation
to discuss the large $z$ behavior.

Now we consider the boundary contribution under the primary
deformation $\Spab{1|n}$. First by our above analysis, vertices of
$L_{(3)}, L_{(4)}$ will not contribute $z$ factors. The
$z$-dependence coming from bosonic and fermionic propagators. Since
bosonic propagator is ${1\over z}$ and fermionic propagator is
${1\over z^0}$, the worst Feynman diagrams are these without bosonic
propagators and scale as  ${1\over z^0}$. Now we consider the vertex
$n$ attached. If $1$ is not attached to same vertex, there is one
fermionic propagator depending on $z$ connected to $n$. If it is
Yukawa coupling we will have $ {\ket{P+zq|n}\over
(P+zq)^2}={\ket{P|n}\over (P+zq)^2}$. If it is gauge coupling we
will have $ {\ket{P+zq|\gamma^\mu|n}\over
(P+zq)^2}={\ket{P|\gamma^\mu|n}+ z\ket{n}q^\mu\over (P+zq)^2}$ where
we have used $\ket{q\gamma^\mu|n}=\ket{n}\Spba{1|\gamma^\mu|n}\sim
\ket{n}q^\mu$. However, under the Light-cone gauge, $z\ket{n}q^\mu
\cdot A^\mu=0$. Thus there is an overall ${1\over z}$ contribution
from the vertex $n$ attached and these   Feynman diagrams do not
give boundary contribution.

\begin{figure}[htb]
\centering
\includegraphics[scale=1]{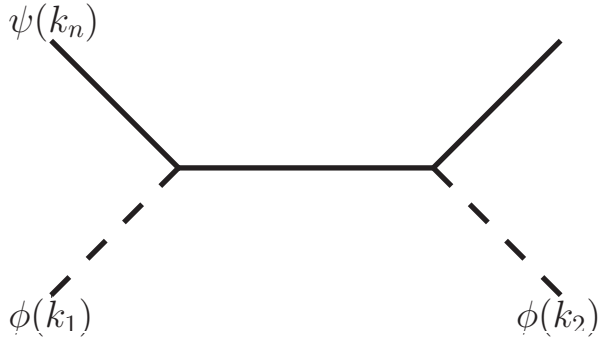}
\caption{The simpliest diagram in
$\langle\phi(k_1)\phi(k_2)|\bar{\psi}(k_n)]$ shift.}
\label{fig:phiphipsibar1}
\end{figure}

By above analysis, we see that boundary contribution comes from
these diagrams $1,n$ attached to same vertex. For these diagrams, we
take the second deformation $\Spab{2|n}$. Similar analysis as above,
only fermionic hard lines (i.e., propagators depending on $z$)
matter. The simplest  diagram of the second shift  is shown in
Figure \ref{fig:phiphipsibar1}. Now let us concentrate to the vertex
$1,n$ attached. Using Feynman rule, we have expression like
\begin{equation}\label{phiphipsibar1vanish}
\frac{(k_1+k_n+z q_i)|n\rangle}{(k_1+k_n+z q_i)^2}=\mathcal{O}(\frac{1}{z})
\end{equation}
Thus by two steps, our algorithm can determine whole amplitude in
the case (c).

%

%%%%%%%%%%%%%%%%%%%%%%%
\subsection{The case (d) with only scalars}
%%%%%%%%%%%%%%%%%%%%%%%

\begin{figure}[htb]
\centering
\includegraphics[scale=0.8]{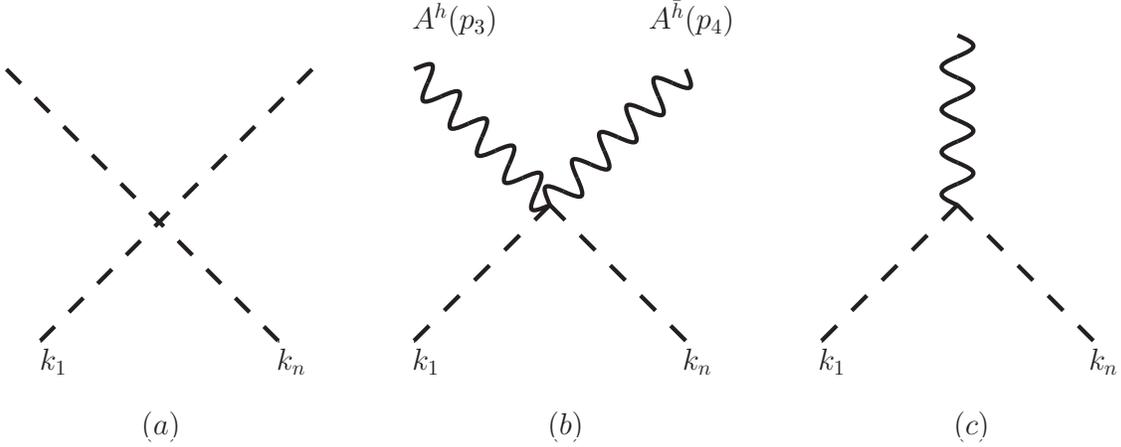}
\caption{Diagrams contributes to boundary terms in a
$\langle\phi(k_1)|\phi(k_n)]$ shift. The third diagram scales as
$\mathcal{O}(z)$, while the other two diagrams scale as
$\mathcal{O}(z^0)$.} \label{fig:scalarshift1}
\end{figure}

Now we consider the large $z$ behavior under the primary deformation
$\Spab{1|n}$. First there is a special diagram (see Figure
\eref{fig:scalarshift1}(c)) for which the Light-cone Lagrangian
\eref{Light-cone} is not well defined. It scales as ${\cal O}(z)$.
For other diagrams, we can use Light-cone Lagrangian
\eref{Light-cone} to analyze. The good point is that all vertices
does not give any $z$ contribution. Furthermore, since all external
particles are scalars, there is no fermionic propagator. Thus we are
left with only bosonic propagators with scale ${1\over z}$. To have
boundary contribution, there should be no any propagator depending
on $z$, thus we are left with only two types of diagrams (see Figure
\eref{fig:scalarshift1}(a),(b)). Now we analyze these three types of
diagrams one by one.

\begin{figure}[htb]
\centering
\includegraphics[scale=0.8]{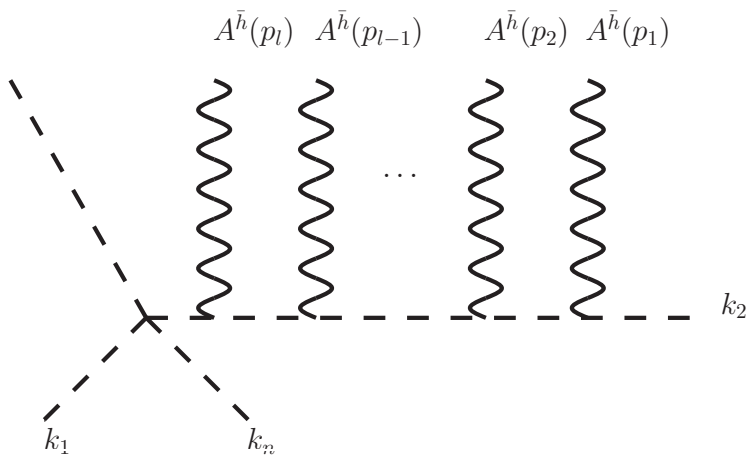}
\caption{A $\mathcal{O}(z^0)$ diagram in a $\langle\phi(k_1)\phi(k_2)|\phi(k_n)]$
shift.} \label{fig:scalarshift32}
\end{figure}

~\\ {\bf Type (a):} First let us consider the second deformation
$\Spab{2|n}$. Under this deformation, only $\partial^{\O h}$
operator contributes $z$  in \eref{Light-cone}. Thus when combining
${1\over z}$ contributions from each bosonic propagator, only type
given in Figure \eref{fig:scalarshift32} has ${\cal O}(1)$ large $z$
behavior. In particular, the hard line (i.e., the $zq$ floating
line) can not have gluon propagator. Otherwise, it will be at least
the ${1\over z}$ scaling. For this type of diagrams, the
$z$-dependent part can be written down as
\begin{equation}
\frac{(2k_2+p_1)\cdot\bar{e}(p_1)(2k_2+2p_1+p_2)\cdot\bar{e}(p_2)\cdots
(2k_2+2p_1+\cdots
2p_{l-1}+p_l)\cdot\bar{e}(p_l)}{2^{\frac{l}{2}}(k_2+p_1)^2(k_2+p_1+p_2)^2\cdots
(k_2+p_1+\cdots p_l)^2}
\end{equation}
where $k_2,k_n$ are shifted and $\O e(p)=\O h-{\O h\cdot p\over
q_1\cdot p}$ (and similarly $ e(p)= h-{ h\cdot p\over
q_1\cdot p}$)~. From it, we can read out boundary of the
deformation $\Spab{2|n}$ as
\begin{equation}
\frac{(q_2\cdot k_1)^l}{2^{\frac{l}{2}}q_2\cdot p_1
q_2\cdot(p_1+p_2)^2\cdots q_2\cdot (p_1+\cdots
p_l)^2}~~~\label{4phi-22}
\end{equation}
multiplying by other factors from remaining part of Feynman
diagrams.

\begin{figure}[htb]
\centering
\includegraphics[scale=0.8]{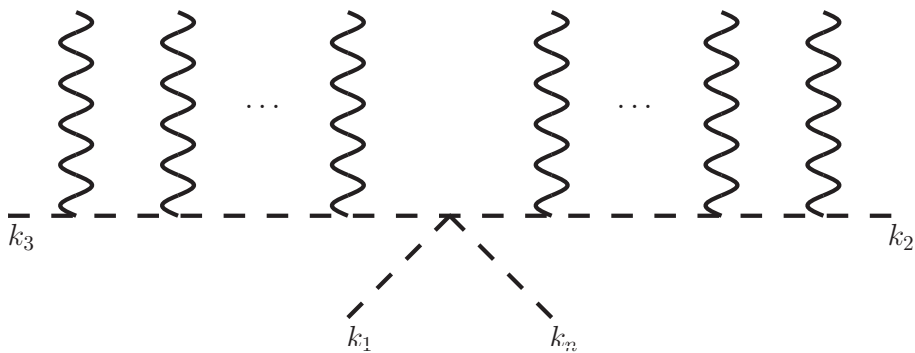}
\caption{A  $\mathcal{O}(z^0)$ diagram in a
$\langle\phi(k_1)\phi(k_2)\phi(k_3)|\phi(k_n)]$ shift.}
\label{fig:scalarshift33}
\end{figure}

For the third deformation $\Spab{3|n}$, by similar analysis,
especially there is no gluon propagator along the hard line,  only
the type of diagrams in Figure \eref{fig:scalarshift33} scales as
${\cal O}(1)$. It is worth to mention that factor \eref{4phi-22} is
not affected by the deformation $\Spab{3|n}$ although other part of
Feynman diagrams will be affected in general.

Finally for the fourth deformation $\Spab{4|n}$, by similar
analysis, especially when the hard line has gluon propagators
scaling behavior will be suppressed by an extra ${1\over z}$ factor,
we found that no matter how we insert the particle $4$ into Figure
\ref{fig:scalarshift33}, we will always get at least ${1\over z}$
scaling. Thus by three steps, we can completely determine boundary
contributions of Figure \eref{fig:scalarshift1}(a).

\begin{figure}[htb]
\centering
\includegraphics[scale=0.7]{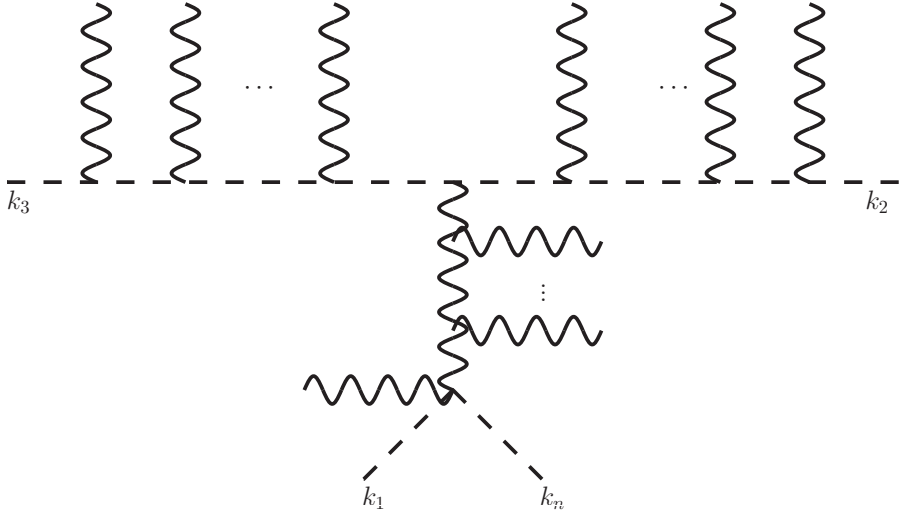}
\caption{A $\mathcal{O}(z^0)$ diagram in a
$\langle\phi(k_1)\phi(k_2)\phi(k_3)|\phi(k_n)]$ shift.}
\label{fig:scalarshift23}
\end{figure}

~\\{\bf Type (b):} Now we consider the second deformation
$\Spab{2|n}$ for the type (b) in Figure \ref{fig:scalarshift1}.
Unlike the type (a) where $1,n$ are attached to $\phi^4$ vertex,
here $1,n$ are attached to $\phi^2 A^h A^{\O h}$ vertex. If the
particle $2$ is along the line $A^{\O h}$, it can be shown using
Lagrangian \eref{Light-cone} that the large $z$ behavior is ${1\over
z}$ at least. But if particle $2$ is along the line $A^{ h}$, it
will contribute to boundary part. After this we will get diagrams
like these given in Figure \ref{fig:scalarshift23}. Next we consider
the third  deformation $\Spab{3|n}$. There are two cases. For the
first case $3$ is not directly connected to $2$ by scalar line, thus
using the same analysis for the type (a), it is ${1\over z}$
behavior at least. For the second case, $3$ is  directly connected
to $2$ by scalar line, thus like the Figure \ref{fig:scalarshift33},
it gives nonzero boundary contributions. Finally, like the case (a),
the fourth deformation $\Spab{4|n}$ will make diagrams in
\ref{fig:scalarshift23} vanishing at large $z$ limit.

\begin{figure}[htb]
\centering
\includegraphics[scale=0.5]{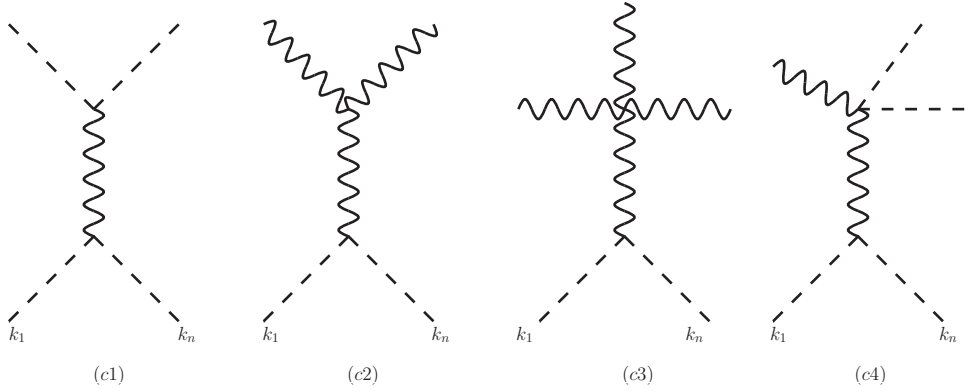}
\caption{Four different diagrams containing the $A\phi\phi$ vertex with $q\cdot P=0$ for the gluon.}
~~~\label{fig:scalarshiftepsilon}
\end{figure}

~\\ {\bf Type (c):} The type (c) of Figure \ref{fig:scalarshift1} is
most complicated one because under our light-cone gauge choice,
$p_{1n}\cdot q=0$, thus we can not impose Light-cone gauge on the
gauge field $A$. To solve this problem, we shift momentum basis to
\begin{equation}
\begin{aligned}
q^{\mu}_{\epsilon}=&\frac{1}{2}[1|\gamma^{\mu}|n_{\epsilon}\rangle,\
\bar{q}^{\mu}_{\epsilon}=\frac{[n|\gamma^{\mu}|1\rangle}{2k_{1}\cdot
k_{n\epsilon}},~~
h^{\mu}_{\epsilon}=&\frac{k_{n\epsilon}^{\mu}}{k_{1}\cdot
k_{n\epsilon}},\
\bar{h}^{\mu}_{\epsilon}=k_1\\
\end{aligned}
\end{equation}
where $|n_{\epsilon}\rangle=|n]+\epsilon|y\rangle$. Thus when we
take $\eps\to 0$ after finishing calculations, we will come back to
original light-cone gauge. The type (c) can grow to following four
diagrams given in Figure \ref{fig:scalarshiftepsilon}. Now we
discuss these four diagrams one by one.

For diagram (c1), using Feynman rules given in \eref{Light-cone}, it is easy to find
\begin{equation}
\begin{aligned}
\frac{1}{\sqrt{2}}(k_1-k_n)\cdot \bar{e}_{\epsilon}=&\frac{1}{\sqrt{2}}(k_1-k_n)\cdot  (\bar{h}_{\epsilon}-\frac{(k_1+k_n)\cdot \bar{h}_{\epsilon}}{(k_1+k_n)\cdot q_{\epsilon}}q_{\epsilon})=0\\
\frac{1}{\sqrt{2}}(k_1-k_n)\cdot e_{\epsilon}=&\frac{1}{\sqrt{2}}(k_1-k_n)\cdot  (h_{\epsilon}-\frac{(k_1+k_n)\cdot h_{\epsilon}}{(k_1+k_n)\cdot q_{\epsilon}}q_{\epsilon})=\sqrt{2}\\
\end{aligned}
\end{equation}
thus If the bottom vertex is $MHV$, the diagram vanishes. If the
bottom vertex is $\overline{MHV}$, the diagram reads
\begin{equation}
\frac{1}{(k_1+k_n)^2}(p_5-p_6)\cdot
\bar{e}_{\epsilon}=\frac{1}{2k_1\cdot k_n}(p_5-p_6)\cdot
(\bar{h}_{\epsilon}-\frac{k_n\cdot \bar{h}_{\epsilon}}{k_n\cdot
q_{\epsilon}}q_{\epsilon})
\end{equation}
This quantity goes to infinity when $\epsilon\rightarrow 0$.
However, for  four scalars, term $({1\over \partial^-}{\cal J}^-)^2$
in the $L_{(4)}$ part of Lagrangian will be singular too. Its
contribution is
\begin{equation}
\begin{aligned}
&-\frac{1}{2}\frac{(k_1-k_n)\cdot q_{\epsilon}(p_5-p_6)\cdot q_{\epsilon}}{[(k_1+k_n)\cdot q_{\epsilon}]^2}\\
\end{aligned}
\end{equation}
When combining these two together, we arrive
\begin{equation}
\begin{aligned}
&\frac{1}{2k_1\cdot k_n}(p_5-p_6)\cdot (\bar{h}_{\epsilon}-\frac{k_n\cdot \bar{h}_{\epsilon}}{k_n\cdot q_{\epsilon}}q_{\epsilon})-\frac{1}{2}\frac{(k_1-k_n)\cdot q_{\epsilon}(p_5-p_6)\cdot q_{\epsilon}}{[(k_1+k_n)\cdot q_{\epsilon}]^2}
=\frac{k_1\cdot (p_5-p_6)}{2k_1\cdot k_n}\\
\end{aligned}\label{phi4limit}
\end{equation}
where the $\eps$ has been canceled out.

It is worth to notice that  \eqref{phi4limit} scales as
$\mathcal{O}(z)$ for $\Spab{1|n}$-deformation, but under
$\Spab{i|n}$-deformation  it scales as  $\mathcal{O}(1)$ ( notice
that $p_5$ or $p_6$ will be shifted). Thus after two steps of
$\Spab{1|n}$ and $\Spab{2|n}$, we arrive to the similar structure as
Figure \ref{fig:scalarshift32}. Thus by same argument, two further
deformations will enable us to detect all contributions.

For the diagram (c2),  the same bottom vertex  must be
$\overline{MHV}$  to give nonzero contribution. If the top vertex is
$\overline{MHV}$, it reads
$(\frac{p_5^h}{p_5^-}-\frac{p_6^h}{p_6^-})p^-$, thus
$p^-=q_{\epsilon}\cdot p$ vanishes when $\epsilon\rightarrow 0$. For
the top vertex to be $MHV$, we choose the leg $p_5$ to have positive
helicity, then the diagram gives (including the contribution of
4-vertex) $-\frac{\sqrt2 k_1\cdot p_6}{(k_1+k_n)^2}
+\frac{1}{2\sqrt2}+\mathcal{O}(\epsilon)$. In fact, this calculation
shows that the diagram (c2) is similar to the type (b) of Figure
\ref{fig:scalarshift1}. Thus by same argument, we need at most four
steps of deformations to determine its contributions.

For the last diagrams (c3), (c4), to have nonzero contribution, the
bottom vertex must be $\overline{MHV}$, while the top vertex is
$z^0$ scaling. Thus under the second deformation $\Spab{2|n}$, whole
diagram scales as ${1\over z}$. In other words, by two steps of
deformations, we can determine its contributions.

~\\{\bf Conclusion:} After above careful analysis, we can see that
for all external particles to be scalars, at most four steps of
deformations are enough to completely determine amplitudes by our
algorithm.

%%%%%%%%%%%%%%%%%%%%
\section{Examples}
%%%%%%%%%%%%%%%%%%%

In this section, we will use two examples to demonstrate our method.
These two examples correspond  to the case (c) and (d) in previous
section. We will use $A$ and $B$ to denote color ordered amplitudes and boundary contributions, and  $\mathcal{A}$ and $\mathcal{B}$ to denote the complete amplitudes and boundary contributions dressed with color factors.

%%%%%%%%%%%%%%%%%%%%
\subsection{Example I: two fermions with three scalars}
%%%%%%%%%%%%%%%%%%%

The first example we will consider is
$\mathcal{A}\left(\bar{\phi}_3(k_1)\phi^1(k_2)\bar{\phi}_1(k_3)
\bar{\psi}_2(k_4)\bar{\psi}_1(k_5)\right)$ in $\mathcal{N}=4$ SYM.
The two fermions are different flavors and complex scalars are
$\phi^i=\phi^{i4}$. In fact, in the language of fermionic
coordinator, their types are given by
$\eta_1^{12}\eta_2^{14}\eta_3^{23}\eta_4^{134}\eta_5^{234}$, so
$1,2,3,4$ appear three times in the superscript. From the known
result of ${\cal N}=4$ SYM theory, we can read out the expression
directly as
\bea \mathcal{A}(12345) &= & \sum_{\sigma\in
S_3(\{2,3,4\})}A(1\sigma_2\sigma_3\sigma_4 5)F^{a_1a_{\sigma_2}a_{\sigma_3}a_{\sigma_4}a_5},\nn
A(1\sigma_2\sigma_3\sigma_4 5) & = &
{-[12][13][24][35]\over[1\sigma_2][\sigma_2\sigma_3][\sigma_3\sigma_4][\sigma_4 5][51]}~~~\label{Exa1-1}
\eea
where we have defined the color factor
\bea F^{a_1a_2a_3a_4a_5}=f^{a_1a_2b}f^{ba_3c}f^{ca_4a_5} \eea
For this example, we need two steps to determine the amplitude
according to our discussion in case (c) of section four.

We will start with deformation $\Spab{1|5}$. Under this
shift, the recursive part gives
\bea & &
\mathcal{A}(\hat{1},3,4,\hat{p}_{25})\frac{1}{p_{25}^2}\mathcal{A}(-\hat{p}_{25},2,\hat{5})+
\mathcal{A}(\hat{1},2,3,\hat{p}_{45})\frac{1}{p_{45}^2}\mathcal{A}(-\hat{p}_{45},4,\hat{5})\nn
& = & \left(
\frac{[23]}{[25][34]}F^{a_1a_3a_4a_2a_5}-\frac{[13][24]}{[14][25][34]}F^{a_1a_4a_3a_2a_5}\right)+\left(
\frac{1}{[45]}F^{a_1a_3a_2a_4a_5}-\frac{[13][24]}{[14][23][45]}F^{a_2a_3a_1a_4a_5}\right)
~~~\label{Exa-15-1}\eea
Now we calculate the boundary part using the boundary recursion
relation with deformation $\Spab{2|5}$. Naively, there will be
following splitting diagrams: (1) ${\cal A}_4(2,3,4,P) {\cal
B}_3^{\langle 1|5]}(-P,1,5)$, (2) ${\cal A}_3(2,3,P) {\cal
B}_4^{\langle 1|5]}(-P,1,5,4)$ and (3) ${\cal A}_3(2,4,P) {\cal
B}_4^{\langle 1|5]}(-P,1,5,3)$. Among these three cases, only case (1)
gives nonzero contribution. The reason is that the sum $k$ of
negative helicity should be four (i.e., we should have $\eta^A$ with
$A=1,2,3,4$ appear four times), thus since one side is four-point
amplitude with $k=2$, another side of three-point amplitude must
have $k=2$. This can not be true for ${\cal A}_3(2,4,P)$ with
$\eta_2^{14}\eta_3^{134}\eta_P^{i}$ or
$\eta_2^{14}\eta_3^{134}\eta_P^{ijk}$ no matter how we choose $i\neq
j\neq k$ from $\{1,2,3,4\}$. For ${\cal A}_3(2,3,P)$, we need to
choose $\eta_2^{14}\eta_3^{23}\eta_P^{1234}$, but since now ${\cal
A}_3$ is MHV amplitude and  the deformation of $\la_2$ makes
$\WH\la_2\sim \la_3\sim \WH \la_P$, we get zero.

For the remaining case (1), we calculate as following. First from
\bea
\mathcal{A}(\bar{\psi}_2(k_6),\bar{\phi}_3(k_1),\bar{\psi}_1(k_5))=-\langle56\rangle
f^{a_1a_5a_6}\eea
we find the boundary part
\begin{equation}
{\cal B}_3^{\langle 1|5]}(-P,1,5)=-\langle 5 P] f^{a_Pa_1a_5}
\end{equation}
Putting it back we get
\bea \mathcal{B}^{\langle 1|5]}&
=&\mathcal{A}_4(\hat{2},3,4,\hat{p}_{15})\frac{1}{p_{15}^2} {\cal
B}_3^{\langle 1|5]}(-\hat{p}_{15},1,\hat{5})\nn
&= &
\frac{[12][13]}{[14][15][23]}F^{a_1a_5a_4a_2a_3}-\frac{[13]}{[15][34]}F^{a_1a_5a_2a_3a_4}
~~~\label{Exa-15-2} \eea
It is easy to check that combining \eref{Exa-15-2} and
\eref{Exa-15-1}, we do reproduce \eref{Exa1-1}.

%%%%%%%%%%%%%%%%%%%%
\subsection{Example II: six scalars}
%%%%%%%%%%%%%%%%%%%

The theory we are considering is the scalar-Yang-Mills theory
\begin{equation}
L=\Tr\left(-\frac{1}{4}F_{\mu\nu}F^{\mu\nu}-D_{\mu}\bar{\Phi}
D^{\mu}\Phi-\frac{g^2}{2}[\Phi,\bar{\Phi}]^2\right)
\end{equation}
For this theory, a standard method is to consider the color-ordering
amplitudes. For six-point amplitudes with three $+$ scalars and
three $-$ scalars, there are following three primary color ordering
amplitudes
\begin{equation}
\begin{aligned}
A(+++---)=&-\frac{\langle12\rangle[45][3|4+5|6\rangle}{\tau_{345}
\langle16\rangle[34][5|3+4|2\rangle}+\frac{\langle23\rangle[56][1|5+6|4\rangle}
{\tau_{234}\langle34\rangle[16][5|3+4|2\rangle}~~~~\label{Exa2-6-1}
\end{aligned}
\end{equation}
\begin{equation}
\begin{aligned}
A(++-+--)=&-\frac{[3|1+2|4\rangle^2\langle56\rangle[12]}
{\tau_{123}[1|2+3|4\rangle[3|4+5|6\rangle\langle45\rangle[23]}
-\frac{[1|2+4|3\rangle^2\langle24\rangle^2[56]}
{\tau_{234}[5|3+4|2\rangle[1|2+3|4\rangle\langle23\rangle\langle34\rangle[16]}\\
&-\frac{[4|3+5|6\rangle^2\langle12\rangle[35]^2}
{\tau_{345}[5|3+4|2\rangle[3|4+5|6\rangle\langle16\rangle[34][45]}~~~~\label{Exa2-6-2}
\end{aligned}
\end{equation}
\begin{equation}
\begin{aligned}
A(+&-+-+-)=-\frac{[2|4+6|5\rangle^2\langle46\rangle^2[13]^2}
{\tau_{123}[1|2+3|4\rangle[3|4+5|6\rangle\langle45\rangle\langle56\rangle[12][23]}\\
&-\frac{[6|1+5|3\rangle^2\langle24\rangle^2[15]^2}
{\tau_{234}[1|2+3|4\rangle[5|3+4|2\rangle\langle23\rangle\langle34\rangle[56][16]}
-\frac{[4|3+5|1\rangle^2\langle26\rangle^2[35]^2}
{\tau_{345}[5|3+4|2\rangle[3|4+5|6\rangle\langle12\rangle\langle16\rangle[34][45]}~~~~\label{Exa2-6-3}
\end{aligned}
\end{equation}
We will calculate \eref{Exa2-6-3} using our algorithm.

We start with deformation $\Spab{1|6}$, the recursive part
is given by
\begin{equation}
\begin{aligned}
\mathcal{A}^{\langle 1|6]}
=&-\frac{\langle16\rangle[35]^2[4|1+6|2\rangle^2}{\tau_{612}\langle12\rangle[34][45]
[5|1+6|2\rangle[3|1+2|6\rangle}
+\frac{[16]\langle24\rangle^2[5|1+6|3\rangle^2}{\tau_{234}[56]\langle23\rangle
\langle34\rangle[1|2+3|4\rangle[5|1+6|2\rangle}\\
&+\frac{[13]^2\langle46\rangle^2[1|2+3|5\rangle^2[2|1+3|6\rangle^2}{\tau_{123}[12][23]
\langle45\rangle\langle56\rangle[1|2+3|4\rangle[3|1+2|6\rangle[1|2+3|6\rangle^2}\\
\end{aligned}\label{Exa2-rec-1}
\end{equation}
We will use another three deformations to  detect the boundary part. In the first step, using
\bea {\cal B}^{\langle1|6]}(g^-(k_7),4,5,6,1) &= &
\frac{[14]\Bigl(-2[15][46]+[14][56]\Bigr)}{[16][17][45][47]}\nn
{\cal B}^{\langle1|6]}(\bar{\Phi}(k_7),5,6,1)&= &
\frac{-\langle17\rangle\langle56\rangle-
2\langle15\rangle\langle67\rangle}{\langle16\rangle\langle57\rangle}\nn
{\cal B}^{\langle1|6]}(g^-(k_7),6,1) & = &
\frac{\langle17\rangle\langle67\rangle}{\langle16\rangle} \eea
 we find
\bea  {\cal A}^{\langle12|6]} &
=&A(\hat{2},3,-\hat{p}_{23})\frac{1}{p_{23}^2}{\cal
B}^{\langle1|6]}(\hat{p}_{23},4,5,\hat{6},1)
+A(\hat{2},3,4,-\hat{p}_{234})\frac{1}{p_{234}^2}{\cal
B}^{\langle1|6]}(\hat{p}_{234},5,\hat{6},1)\nn
& &+A(\hat{2},3,4,5,\hat{p}_{16})\frac{1}{p_{16}^2}{\cal
B}^{\langle1|6]}(-\hat{p}_{16},\hat{6},1)\nn
&
=&\frac{[14]\langle26\rangle\langle36\rangle\Bigl([14][5|2+6|3\rangle-
2[15][4|2+6|3\rangle\Bigr)}{[45]\langle23\rangle[1|2+6|3\rangle[1|2+3|6\rangle[4|2+3|6\rangle}\nn
&+&
\frac{[24]^2[3|2+4|6\rangle^2\Bigl(2\langle15\rangle[2|3+4|6\rangle+\langle56\rangle
[2|3+4|1\rangle\Bigr)}{\tau_{234}[23][34]\langle16\rangle[2|3+4|5\rangle[2|3+4|6\rangle[4|2+3|6
\rangle}\nn
&+&
\frac{[26]\langle35\rangle^2[1|2+6|4\rangle^2}{\tau_{345}[16]\langle34\rangle\langle45
\rangle[2|3+4|5\rangle[1|2+6|3\rangle}~~~~~\label{Exa2-boun-1} \eea
In the second step, using
\begin{equation}
\begin{aligned}
{\cal B}^{\langle12|6]}(g^-(k_7),5,6,1,2)&=\frac{[25]}{[27][57]},~~~~
{\cal B}^{\langle12|6]}(\Phi(k_7),6,1,2)&=-1\\
\end{aligned}
\end{equation}
we find
\bea {\cal A}^{\langle123|6]}&
=&A(\hat{3},4,-\hat{p}_{34})\frac{1}{p_{34}^2}{\cal B}^{\langle12|6]}
(\hat{p}_{34},5,\hat{6},1,2)
+A(\hat{3},4,5,-\hat{p}_{345})\frac{1}{p_{345}^2}{\cal
B}^{\langle12|6]}(\hat{p}_{345},\hat{6},1,2)\nn
&=&-\frac{[25]\langle36\rangle\langle46\rangle}{\langle34\rangle[2|3+4|6\rangle[5|3+4|6\rangle}
+\frac{[35]^2[4|3+5|6\rangle^2}{\tau_{345}[34][45][3|4+5|6\rangle[5|3+4|6\rangle}
~~~~~\label{Exa2-boun-2} \eea
In the third step with deformation,  using
\begin{equation}
\begin{aligned}
{\cal B}^{\langle123|6]}(g^-(k_7),6,1,2,3)&=\frac{[13]^2[27]}{[23][37][17]^2}\\
\end{aligned}
\end{equation}
we find
\bea {\cal
A}^{\langle1234|6]}&=&A(\hat{4},5,-\hat{p}_{45})\frac{1}{p_{45}^2}{\cal
B}^{\langle123|6]}(\hat{p}_{45}, \hat{6},1,2,3)\nn
&=&\frac{[13]^2[46]\langle56\rangle[2|4+5|6\rangle}{[23]\langle45\rangle[3|4+5|6
\rangle[1|4+5|6\rangle^2}~~~~~\label{Exa2-boun-3} \eea

Numerical checking shows 
$\mathcal{A}=\mathcal{A}^{\langle 1|6]}+\mathcal{A}^{\langle 12|6]}+\mathcal{A}^{\langle 123|6]}+\mathcal{A}^{\langle 1234|6]}$ , although it is a
little bit complicated  to show it analytically.

%%%%%%%%%%%%%%%%%%%%%%%%%%%
\section{Discussions}
%%%%%%%%%%%%%%%%%%%%%%%%%%%%

In this paper we showed that boundary contributions satisfy similar recursion relations as scattering amplitudes, and presented a new algorithm to compute boundary contributions. We analyzed large $z$ scaling of amplitudes and boundary contributions in standard model like theories via light cone gauge, and gave two explicit calculations.

%We analyzed large $z$ scaling of amplitudes and boundary contributions via light cone gauge, and proved in standard model like theories, the complete amplitude can be computed using (at most) 4 step recursions.

It is worth noting that although we only discussed on-shell amplitudes, our method can be applied to amplitudes with off shell currents. The recursion relations for amplitudes with off shell currents was discussed in \cite{Feng:2011twa}, and one complication there was besides physical states $g^{\pm}$, longitudinal and time-like states also contributes. Fortunately, only physical states contributes in light cone gauge.

In section 2, we mentioned in general two deformations do not commute. It would be interesting to investigate what is the commutator of two deformations. And it might help us to determine the best choice of deformations.

In general, with more derivatives in vertices the large $z$ scaling of amplitudes get worse. Thus for  these theories  the boundary might not vanish when all possible deformations of the type $\Spab{i|n}$ are exploited. It would be interesting to decide under which conditions, our new
 recursion algorithm ends in finite steps. One particular theory is the one with matters coupling to gravity. It would be nice if we can give a similar analysis for the theory, like the example (i.e., the
 standard model like theory) studied in the paper.

%From section 4 we can see it is a hard task to find good shifts by analyzing off shell diagrams, even in light cone gauge. In a coming paper, we will present an on shell method to determine good deformations.

Last, let us point out that  boundary contributions serves as a bridge between on-shell and off-shell quantities. On one hand, boundary contributions stem from on shell scattering amplitude, and can be computed using on-shell methods. On the other hand, in many cases Feynman diagrams contributing to boundary part $B^{\langle1|n]}$ have the topology of the vertex $V(1,n,P)$ connected to off-shell current $J(23\cdots n-1,P)$ (where $P$ is the propagator). So using boundary contributions we can compute off-shell quantities like correlations function effectively.

%%%%%%%%%%%%%%%%%%%%%%%%%
\section*{Acknowledgement}
%%%%%%%%%%%%%%%%%%%%%%%%

We thank Radu Roiban, Chenkai Qiao, Junjie Rao and
Kang Zhou  for helpful discussions. We would also like to thank Rijun Huang, Congkao Wen for pointing out typos in a previous version. This work is supported by
Qiu-Shi funding and Chinese NSF funding under contracts No.11031005,
No.11135006, No.11125523 and No.10875103, and National Basic
Research Program of China (2010CB833000).

%%%%%%%%%%%%%%%%%%%%%%%%%%%%%%%%%%%%%%%%%%%%%%%%%%%%%%
\appendix
%%%%%%%%%%%%%%%%%%%%%%%%%%%%%%%%%%%%%%%%%%%%%%%%%%%

%%%%%%%%%%%%%%%%%%%%%%%%%%%
\section{Ordering of integration and residues of multiple variables\label{appendixA}}
%%%%%%%%%%%%%%%%%%%%%%%%%

In this part, we discuss some aspects of integration of multiple variables related to our study.

First we recall the {\bf Fubini-Tonelli theorem} (often just called Fubini's theorem) which states that if $X$ and $Y$  are $\sigma$-finite measure spaces, and if $f$ is a measurable function such that any one of the three integrals
\bea (I)=\int_X \left( \int_Y |f(x,y)| dy\right)dx,~~~(II)=\int_Y \left( \int_X |f(x,y)| dx\right)dy,
~~~(III)=\int_{X\times Y} |f(x,y)|d(x,y)\eea
is finite, then
\bea \int_X \left( \int_Y f(x,y) dy\right)dx=\int_Y \left( \int_X f(x,y) dx\right)dy=\int_{X\times Y} f(x,y)d(x,y)~~~~\label{Fubini}\eea
In other words, the ordering of two integrations can be exchanged. This theorem is very important
for our derivation of recursion relation of boundary contributions.

Having known when the ordering of integrations can be exchanged,  we present a counter  example related to our discussion in the paper. The  example is following two integrations
\bea I_1 & = & \oint_{z_1=0} {dz_1\over z_1} \oint_{z_2=0} {dz_2\over z_2} { z_1+a z_2\over z_1+b z_2},
~~~~~I_2  =  \oint_{z_2=0} {dz_2\over z_2}\oint_{z_1=0} {dz_1\over z_1}  { z_1+a z_2\over z_1+b z_2}~~~
\label{Counter-1}\eea
It is easy to see, depending on the value of $z_1$ we have
\bea I_{12}(z_1)\equiv \oint_{z_2=0} {dz_2\over z_2} { z_1+a z_2\over z_1+b z_2}=\left\{\begin{array}{ll}
1,~~~~ & z_1\neq 0 \\  {a\over b},~~~ & z_1=0 \end{array}\right.~~~
\label{Counter-1-0}\eea
thus we have $I_1=\oint_{z_1=0} {dz_1\over z_1} I_{12}(z_1) = 1$ since no matter
how small is the circle around $z_1=0$, as long as $z_1\neq 0$, $I_{12}(z_1\neq 0)= 1$. Similarly $I_2=\oint_{z_2=0}
{dz_2\over z_2} I_{21}(z_2)=\oint_{z_2=0}
{dz_2\over z_2}  {a\over b}=  {a\over b}$. Thus we see that $I_1\neq
I_2$, i.e., the ordering of contour integrations can not be exchanged.

The  reason of non-commutativity of \eref{Counter-1} is not exactly the one
mentioned in  the { Fubini-Tonelli theorem}. When we parameterize $z_1=R_1
e^{i\theta_1}$ and $z_2=R_2 e^{i\theta_2}$  with small radius $R_1,
R_2$ for $I_1$, we need to impose condition $R_2<R_1$ to make sure
when we evaluate $z_2$-integration, only $z_2=0$ pole is inside the
circle, i.e., it will not contain pole from factor $z_1+b z_2$.
 Similarly, for $I_2$ we need to have $R_2> R_1$. Thus when we exchange the ordering,
integral regions are, in fact, different.

Example \eref{Counter-1} is, in fact, one example of residues of
multi-variable studied in \cite{ArkaniHamed:2009dn}. Naively exchanging the ordering of integral
variables, the residue can be different only up to a sign. However, as pointed out
in \cite{Sogaard:2014ila}, when the integration is degenerated, we must be careful.
It is easy to see that our example belongs to this special case. If we make following
transformation of variables $f_1= z_1 (z_1+b z_2)$ and $f_2=z_2$,  the Jacobi
${\partial(f_1, f_2)\over \partial(z_1,z_2)}=2z_1+b z_1$ which is
zero when $z_1=z_2=0$.

Now we consider our example using the method given in \cite{Sogaard:2014ila}.
There are three factors $z_1, z_2, z_1+b z_2$ in denominator, thus there are
several possible combinations. For the first combination, we define
$f_1= z_1 (z_1+b z_2)$ and $f_2=z_2$, thus using the algorithm in \cite{Sogaard:2014ila}
we find $h_1=z_1^2$, $h_2=z_2$ and the transformation matrix is given by
\bea \left( \begin{array}{c} h_1 \\ h_2 \end{array}\right)= \left(
\begin{array}{cc} 1 & -b z_1  \\ 0 & 1 \end{array}\right)\left(
\begin{array}{c} f_1 \\ f_2 \end{array}\right)\equiv A \left(
\begin{array}{c} f_1 \\ f_2 \end{array}\right)\eea
Now the integration becomes
\bea \oint dz_1 dz_2 { P(z_1, z_2)\over  f_1 f_2}=\oint {dz_1\over
h_1} \oint {dz_2\over h_2} P(z_1, z_2) {\rm det}(A),\eea
Putting $P(z_1,z_2)=z_1+ a z_2$ back we get
\bea \oint {dz_1\over z_1^2} \oint {dz_2\over z_2} (z_1+a z_2)
=\oint {dz_1 z_1\over z_1^2} \oint {dz_2\over z_2} + \oint
{dz_1\over z_1^2} \oint {dz_2 a z_2\over z_2} =1 \eea
For the second combination $\W f_1= z_1$ and $\W f_2=z_2
(z_1+b z_2)$, we can find $\W h_1=z_1, \W h_2= z_2^2$ and the transformation matrix
is given by
\bea \left( \begin{array}{c} \W h_1 \\ \W h_2 \end{array}\right)=
\left( \begin{array}{cc} 1 & 0  \\ {-z_2\over b} & {1\over b}
\end{array}\right)\left( \begin{array}{c}\W f_1 \\\W f_2
\end{array}\right)\eea
thus the integration becomes
\bea\oint {dz_1\over z_1} \oint {dz_2\over z_2^2} {(z_1+a z_2)\over
b} = \oint {dz_1\over z_1} {z_1\over b}\oint {dz_2\over z_2^2}+\oint {dz_1\over
z_1} \oint {dz_2\over z_2^2} {a z_2\over b}={a\over b}\eea
Finally for the third combination $\WH f_1= z_1 z_2$, $\WH f_2=z_1+b z_2$,  we find $\WH h_1=z_1^2,\WH h_2=z_2^2$, thus
\bea \left( \begin{array}{c} \WH h_1 \\ \WH h_2 \end{array}\right)=
\left( \begin{array}{cc} -b & z_1  \\ {-1\over b} & {z_2\over b}
\end{array}\right)\left( \begin{array}{c}\WH f_1 \\\WH f_2
\end{array}\right)\eea
So the integration becomes
\bea\oint {dz_1\over z_1^2} \oint {dz_2\over z_2^2} (z_1+a z_2)z_2
 = \oint {dz_1 z_1\over z_1^2} \oint {dz_2 z_2\over z_2^2}+\oint {dz_1\over z_1^2} \oint {dz_2 a z_2^2\over z_2^2}=1\eea

From above calculations,  we see that different orderings of integrations in \eref{Counter-1} correspond to different combinations from the point of view of residue of multi-variables.

%%%%%%%%%%%%%%%%%%%%
\section{Other deformations for boundary}
%%%%%%%%%%%%%%%%%%%%%

In the section two, we have written down recursion relation for boundary contributions under the primary
deformation $\Spab{1|n}$ using second deformation $\Spab{2|n}$ ( or further deformations of the type
$\Spab{i|n}$). In this Appendix, we discuss possible recursion relation using other types of deformations.
Before doing so, let us fix the notation that under the primary deformation $\und{0}\equiv\Spab{1|n}$, the amplitude
can be written as
\bea A= {\cal A}^{\und{0}}+ {\cal B}^{\und{0}}~~~~\label{Other-A-0}\eea
where ${\cal R}^{\und{0}}$ is the recursive part  and ${\cal B}^{\und{0}}$ is the boundary part we are trying to determine. We will consider two kinds of other deformations. The first one is another
BCFW-deformation, for example, $\und{1}\equiv \Spab{2|3}$. The second one is the Risager's deformation
\cite{Risager:2005vk}  defined by
\bea \Spbb{ijk|\eta} \equiv \left\{ \begin{array}{l}
\bket{i(z)} = \bket{i}-z \Spaa{j|k} \eta \\ \bket{j(z)} = \bket{j}-z \Spaa{k|i} \eta \\
\bket{k(z)} = \bket{k}-z \Spaa{i|j}\eta\end{array} \right.~~~~\label{Risager-anti}\eea
or
\bea \Spaa{ijk|\eta} \equiv \left\{ \begin{array}{l}
\ket{i(z)} = \ket{i}-z \Spbb{j|k} \eta \\ \ket{j(z)} = \ket{j}-z \Spbb{k|i} \eta \\
\ket{k(z)} = \ket{k}-z \Spbb{i|j}\eta\end{array} \right.~~~~\label{Risager-spinor}\eea
Now we discuss them one by one.

%%%%%%%%%%%%%%%%%%%
\subsection{Using Risager's deformation}
%%%%%%%%%%%%%%%%%

Since from \eref{BPole-1},  spurious poles after the primary deformation $\Spab{1|n}$ are the type of  $\Spab{n|P|1}$ with $p_1,p_n\not\in P$, we would like to take following two kinds of deformations
$\Spbb{ijk|\W\la_1}$ (so $i,j,k\neq 1$) and $\Spaa{ijk|\la_n}$ (so $i,j,k\neq n$), thus  spurious poles are
not deformed. To make our discussion more explicitly we will consider the deformation $\und{1}\equiv \Spaa{234|\la_n}$, thus following propagators $(p_i+P_J)^2, (p_i+p_j+P_J)^2$ with $i\neq j=2,3,4$ and
$J\subset \{5,6,...,n-1\}$\footnote{We will also use the notation $\O J$ which means that
 $J\bigcup \O J=\{5,6,...,n-1\}$.} will provide poles   under the $\und{1}$-deformation.
The locations of these poles are
\bea z_{i,J} = {(p_i+P_J)^2\over \Spab{n|p_i+P_J|i} \Spbb{j|k}},~~~~z_{ij,J} = -{(p_i+p_j+P_J)^2\over \Spab{n|p_i+p_j+P_J|k} \Spbb{i|j}}~~~~\label{Other-C-pole-loca}\eea
with $\{i,j,k\}$ to be the cyclic ordering of $\{2,3,4\}$.
Using the contour integration $\oint_{\infty} {dz\over z}{\cal B}^{\und{0}}(z)$, we can derive the  recursion relation for boundary part
${\cal B}^{\und{0}}$ as following
\bea {\cal B}^{\und{0}} & = &  {\cal B}^{\und{0}\und{1}}+ \sum_{z_{i,J}}
{A_L( \WH\la_i(z_{i,J}), P_J, \WH P(z_{i,J})) {\cal B}^{\und{0}} (-\WH P(z_{i,J}), \WH\la_j(z_{i,J}), \WH\la_k(z_{i,J}), P_{\O J},
p_1,p_n)\over (p_i+P_J)^2}\nn
& & + \sum_{z_{ij,J}}  {A_L( \WH\la_i(z_{ij,J}),\WH\la_j(z_{ij,J}),  P_J, \WH P(z_{ij,J})) {\cal B}^{\und{0}} (-\WH P(z_{ij,J}), \WH\la_k(z_{ij,J}), P_{\O J},
p_1,p_n)\over (p_i+p_j+ P_J)^2}~~~\label{Risager-bound-recur-1}\eea
Since all poles can be proved by same method, we will give the proof for pole $z_{i,J}$ only. The residue
of pole $z_{i,J}$ is given by
\bea  & & \oint_{z_{i,J}} { d z\over z} {\cal B}^{\und{0}}( \la_i-z \Spbb{j|k} \la_n,
\la_j-z \Spbb{k|i} \la_n, \la_k-z \Spbb{i|j}\la_n)~~~\label{Other-C-1} \eea
where the contour is a small circle around pole $z_{i,J}$. Now we put the expression of  ${\cal B}^{\und{0}}=\oint_{\infty} {d w\over w} A(\la_1-w \la_n, \W\la_n+w \W\la_1)$ back to get
\bea  & & \oint_{z_{i,J}} { d z\over z}
\oint_{w=\infty} {dw\over w} A_n( \la_1-w\la_n,\la_i-z \Spbb{j|k} \la_n,
\la_j-z \Spbb{k|i} \la_n, \la_k-z \Spbb{i|j}\la_n, \W\la_n+w \W\la_1)\nn
& = & \oint_{w=\infty} {dw\over w} \oint_{z_{i,J}} { d z\over z}A_n( \la_1-w\la_n,\la_i-z \Spbb{j|k} \la_n,
\la_j-z \Spbb{k|i} \la_n, \la_k-z \Spbb{i|j}\la_n, \W\la_n+w \W\la_1)\nn
& = & -\oint_{w=\infty} {dw\over w} { A_L (\la_i-z_{i,J} \Spbb{j|k} \la_n,P_J, \WH P) A_R (-\WH P, ...,
\la_1-w\la_n,\la_j-z_{i,J} \Spbb{k|i} \la_n, \la_k-z_{i,J} \Spbb{i|j}\la_n, \W\la_n+w \W\la_1)\over (p_i+P_J)^2}\nn
& = & - { A_L (\la_i-z_{i,J} \Spbb{j|k} \la_n,P_J, \WH P) {\cal B}^{\und{0}} (-\WH P, ...,
\la_1,\la_j-z_{i,J} \Spbb{k|i} \la_n, \la_k-z_{i,J} \Spbb{i|j}\la_n, \W\la_n)\over (p_i+P_J)^2} \eea
where at the second line, we have exchanged the ordering of two contour using the Fubini-Tonelli theorem
while at the fourth line, we have used the fact that in the third line, the variable $w$  appears only
on $A_R$. Thus we have proved the  boundary recursion relation \eref{Risager-bound-recur-1}.

%%%%%%%%%%%%%
\subsection{Using the  deformation $\Spab{2|3}$}
%%%%%%%%%%%%%

First, using \eref{BD-exp}, we can see that possible poles of ${\cal B}^{\und{0}}$ will be followings
\bea   & & P_J^2,~~(p_2+P_J)^2,~~(p_3+P_J)^2,~~(p_2+p_3)^2,~~(p_2+p_3+P_J)^2,\nn
& & \Spab{n|P_J|1}^a,~~\Spab{n|p_2+P_J|1}^a,~~\Spab{n|p_3+P_J|1}^a,~~\Spab{n|p_2+p_3+P_J|1}^a,~~ \Spab{n|p_2+p_3|1}^a~~~\label{Other-B-all-poles} \eea
where $J\subset \{4,5,...,n-1\}$. From \eref{Other-B-all-poles}, we observe a  crucial difference between the deformation of $\Spab{i|n}$ type and the deformation $\Spab{2|3}$ is that spurious poles
$\Spab{n|P|1}$ could  be detected by the deformation $\Spab{2|3}$.
Furthermore,  the power of spurious poles could be  bigger than one. From \eref{Other-B-all-poles},
we find locations of poles are\footnote{It is worth to notice that for spurious poles,
$\Spab{n|p_2+P_J|1}$ is same to the one $\Spab{n|p_3+P_{\O J}|1}$ with $\O J\equiv \{4,5,...,n-1\}- J$. Thus when we sum over spurious poles, we should avoid the double counting.}
\bea z_{2,J} & = & {(p_2+P_J)^2\over \Spab{3|P_J|2}},~~~z_{3,J}  =  -{(p_3+P_J)^2\over \Spab{3|P_J|2}}\nn
z_{2,J;s} & = & {\Spab{n|p_2+P_J|1}\over \Spaa{n|3}\Spbb{2|1}},~~~z_{3,J;s}  = - {\Spab{n|p_3+P_J|1}\over \Spaa{n|3}\Spbb{2|1}},\eea
thus we can write down following expression for ${\cal B}^{\und{0}}$ as
\bea {\cal B}^{\und{0}} & =& {\cal B}^{\und{0}\und{1}}
-\sum_{z_{2,J}} \oint_{z=z_{2,J}} {dz\over z} {\cal B}^{\und{0}} (\la_2-z\la_3, \W\la_3+z\W\la_2)-\sum_{z_{3,J}} \oint_{z=z_{3,J}} {dz\over z} {\cal B}^{\und{0}} (\la_2-z\la_3, \W\la_3+z\W\la_2)\nn
& &-\sum_{z_{2,J;s}} \oint_{z=z_{2,J;s}} {dz\over z} {\cal B}^{\und{0}} (\la_2-z\la_3, \W\la_3+z\W\la_2)-\sum_{z_{3,J;s}} \oint_{z=z_{3,J;s}} {dz\over z} {\cal B}^{\und{0}} (\la_2-z\la_3, \W\la_3+z\W\la_2)~~~\label{Other-B0-23-exp} \eea
where ${\cal B}^{\und{0}\und{1}}=\oint_{z=\infty} {dz\over z} {\cal B}^{\und{0}} (\la_2-z\la_3, \W\la_3+z\W\la_2) $ is the remaining boundary part.

For the contour integration around the pole $z_{2,J}$, we can evaluate as following
\bea  & & \oint_{z_{2,J}} { d z\over z} {\cal B}^{\und{0}}( \la_2-z
\la_3, \W\la_3+z\W\la_2)\nn
& = & \oint_{z_{2,J}} { d z\over z} \oint_{w=\infty} {dw\over w}
A_n(\{\la_1-w\W\la_n,\W\la_1\},\{\la_2-z \la_3,\W\la_2\}, \{\la_3,
\W\la_3+z\W\la_2\},...,p_{n-1},\{\la_n, \W\la_n+w \W\la_1\})\nn
& = & \oint_{w=\infty} {dw\over w}\oint_{z_{2,J}} { d z\over z}
A_n(\{\la_1-w\W\la_n,\W\la_1\},\{\la_2-z \la_3,\W\la_2\}, \{\la_3,
\W\la_3+z\W\la_2\}...,p_{n-1},\{\la_n, \W\la_n+w \W\la_1\})\nn
& = & -\oint_{\infty} {dw\over w} \left\{ {A_L(\{\la_2-z_{2,J} \la_3,\W\la_2\},P_J,  \WH P^h) A_R (-\WH
P^{-h}, P_{\O J},\{\la_1-w\W\la_n,\W\la_1\}, \{\la_n, \W\la_n+w
\W\la_1\},\{\la_3, \W\la_3+z_{2,J}\W\la_2\})\over (p_2+P_J)^2} \right\}\nn
& = & - {A_L(\{\la_2-z_{2,J} \la_3,\W\la_2\},P_J,  \WH P^h) {\cal B}^{\und{0}} (-\WH
P^{-h}, P_{\O J},\{\la_1,\W\la_1\},\{\la_n,\W\la_n\},\{\la_3, \W\la_3+z_{2,J}\W\la_2\})\over (p_2+P_J)^2}
~~~\label{Other-B23-reg-1}\eea
where in the third line we have exchanged ordering of contour
integrations and in the fourth line, we have evaluate $z$-contour integration.
Finally since $w$ appears in $A_R$ part only, we arrive the fifth line.
Similarly we can obtain
\bea & & \oint_{z_{3,J}} { d z\over z} {\cal B}^{\und{0}}( \la_2-z
\la_3, \W\la_3+z\W\la_2)\nn
& = & - {A_L(\{\la_3,\W\la_3+z_{2,J}\W \la_2\},P_J,  \WH P^h) {\cal B}^{\und{0}} (-\WH
P^{-h}, P_{\O J},\{\la_1,\W\la_1\},\{\la_n,\W\la_n\},\{\la_2-z_{2,J} \la_3,\W\la_2\})
\over (p_3+P_J)^2}~~~\label{Other-B23-reg-2}\eea

Now we evaluate the contour integration around the pole $z_{2,J;s}$ similarly
\bea  & & \oint_{z_{2,J;s}} { d z\over z} {\cal B}^{\und{0}}( \la_2-z
\la_3, \W\la_3+z\W\la_2)\nn
& = & \oint_{z_{2,J;s}} { d z\over z} \oint_{w=\infty} {dw\over w}
A_n(\{\la_1-w\W\la_n,\W\la_1\},\{\la_2-z \la_3,\W\la_2\}, \{\la_3,
\W\la_3+z\W\la_2\},...,p_{n-1},\{\la_n, \W\la_n+w \W\la_1\})\nn
& = & \oint_{w=\infty} {dw\over w}\oint_{z_{2,J;s}} { d z\over z}
A_n(\{\la_1-w\W\la_n,\W\la_1\},\{\la_2-z \la_3,\W\la_2\}, \{\la_3,
\W\la_3+z\W\la_2\}...,p_{n-1},\{\la_n, \W\la_n+w \W\la_1\})~~~~\label{Other-B23-z2s-1}\eea
Up to this step, there is nothing particular. However, when we try to
evaluate $\oint_{z_{2,J;s}}$ first with $w$ fixed, new phenomenon happens.
In fact, the pole $z_{2,J;s}$ is the large $w$ limit of following two poles
$z_{1,2,J}  =  { (p_1+p_2+P_J)^2-w\Spab{n|p_1+p_2+P_J|1}\over
\Spab{3|p_1+p_2+P_J|2}-w \Spaa{n|3}\Spbb{2|1}}$ coming from the propagator $(p_1+p_2+P_J)^2$ and
$z_{n,2,J}= { (p_n+p_2+P_J)^2+w\Spab{n|p_n+p_2+P_J|1}\over
\Spab{3|p_n+p_2+P_J|2}+w \Spaa{n|3}\Spbb{2|1}}$ coming from the propagator $(p_n+p_2+P_J)^2$.
Thus \eref{Other-B23-z2s-1} should become to
\bea & & -\oint_{w=\infty} {dw\over w}\left\{  {A_L(
\{\la_1-w\W\la_n,\W\la_1\},\{\la_2-z_{1,2,J} \la_3,\W\la_2\},P_J,
\WH P) A_R(-\WH P, P_{\O J}, \{\la_3,
\W\la_3+z_{1,2,J}\W\la_2\},\{\la_n, \W\la_n+w \W\la_1\})\over
(p_1+p_2+P_J)^2}\right. \nn
& & \left. + {A_L( \{\la_n, \W\la_n+w \W\la_1\},\{\la_2-z_{n,2,J}
\la_3,\W\la_2\},P_{ J}, \WH P) A_R(-\WH P, P_{\O J}, \{\la_3,
\W\la_3+z_{n,2,J}\W\la_2\},\{\la_1-w\W\la_n,\W\la_1\})\over
(p_n+p_2+P_{ J})^2}\right\}~~~~\label{Other-B23-z2s-2}\eea
We can not go further from \eref{Other-B23-z2s-2} since variable $w$ appears
in both $A_L, A_R$. However, although we can not finish the evaluation of $w$-contour
integration, each piece in \eref{Other-B23-z2s-2} depends only on lower point amplitudes.
Thus, in a weak sense, it is also a recursion relation.

Similar evaluation around pole $z_{3,J;s}$ will give
\bea & & -\oint_{w=\infty} {dw\over w}\left\{  {A_L(
\{\la_1-w\W\la_n,\W\la_1\},\{\la_3,
\W\la_3+z_{1,3,J}\W\la_2\},P_J,
\WH P) A_R(-\WH P, P_{\O J}, \{\la_2-z_{1,3,J}
\la_3,\W\la_2\},\{\la_n, \W\la_n+w \W\la_1\})\over
(p_1+p_3+P_J)^2}\right. \nn
& & \left. + {A_L( \{\la_n, \W\la_n+w \W\la_1\},\{\la_3,
\W\la_3+z_{n,3,J}\W\la_2\},P_{ J}, \WH P) A_R(-\WH P, P_{\O J}, \{\la_2-z_{n,3,J}
\la_3,\W\la_2\},\{\la_1-w\W\la_n,\W\la_1\})\over
(p_n+p_3+P_{ J})^2}\right\}~~~~\label{Other-B23-z2s-3}\eea
with
\bea
z_{1,3,J} = - {
(p_1+p_3+P_J)^2-w\Spab{n|p_1+p_3+P_J|1}\over \Spab{3|p_1+p_3+P_J|2}-w
\Spaa{n|3}\Spbb{2|1}},~~~z_{n,3,J} = - {
(p_n+p_3+P_J)^2+w\Spab{n|p_n+p_3+P_J|1}\over \Spab{3|p_n+p_3+P_J|2}+w
\Spaa{n|3}\Spbb{2|1}}~~~~~~\label{Other-B-zw-pole}\eea

Finally, putting \eref{Other-B23-reg-1}, \eref{Other-B23-reg-2}, \eref{Other-B23-z2s-2}, and \eref{Other-B23-z2s-3} back to \eref{Other-B0-23-exp}, we get a "weak recursion relation"
for ${\cal B}^{\und{0}}$ using the deformation $\Spab{2|3}$.

%%%%%%%%%%%%%%%%%%%%
\section{Light cone propagator }
%%%%%%%%%%%%%%%%%%%%%%

Although it is not used explicitly, we like to discuss one aspect of
light-cone propagator given by
\begin{equation}
\Pi_{\mu\nu}=\frac{1}{p^2}\left(\eta_{\mu\nu}-\frac{q_{\mu}p_{\mu}+p_{\mu}q_{\mu}}{q\cdot
p}\right)
\end{equation}
Using the basis $q,\O q, h, \O h$ (see \eref{q-basis}),  it is easy
to rewrite it as
\begin{equation}
\begin{aligned}
&\eta_{\mu\nu}-\frac{q_{\mu}p_{\nu}+p_{\mu}q_{\nu}}{q\cdot
p}=e_{\mu}\bar{e}_{\nu}+e_{\nu}\bar{e}_{\mu}-\frac{p^2}{(q\cdot
p)^2}q_{\mu}q_{\nu}\end{aligned}
\end{equation}
where we have  defined
\begin{equation}
e(p)=h-\frac{h\cdot p}{q\cdot p}q,\
~~~\bar{e}(p)=\bar{h}-\frac{\bar{h}\cdot p}{q\cdot p}q,
\end{equation}
which have been used in main text and  are proportional to the gluon
polarization vectors $\epsilon^+(p)$ and $\epsilon^-(p)$,
respectively. The $p^2$ of the term $\frac{p^2}{(q\cdot
p)^2}q_{\mu}q_{\nu}$ will cancel the denominator of $\Pi_{\mu\nu}$,
thus this term will give an effective $4$-point vertex.

%save ~\\ {\bf The pole structure of the residue $C_J$:}  We know that $C_J$ is given by
%\bea C_J & = & A_L (z_J) A_R(z_J),~~~~z_J={ (P_J+p_1)^2\over \Spab{n|P_J+p_1|1}} \eea
%The possible poles of $A_L$ part are $P_{K\subset J}^2, (p_1+P_{K\subset J})^2$. Under the shifting, $(p_1+P_{K\subset J})^2$ becomes
%\bea(p_1+P_{K})^2\to {\Spab{n|(P_J+p_1)(P_K-P_J)(P_K+p_1)|1}\over \Spab{n|P_J+p_1|1}},~~~~K\subset J \eea
%Similarly the possible poles of $A_R$ part are $P_{K\subset \O J}^2, (p_n+P_{K\subset \O J})^2$ with $J\bigcup \O J={\cal T}$ and  under the shifting, $ (p_n+P_{K\subset \O J})^2$ becomes
%\bea(p_n+P_{K})^2\to {\Spab{n|(P_J+p_1)(P_K+P_J+p_1+p_n)(P_K+p_n)|1}\over \Spab{n|P_J+p_1|1}},~~~~K\subset J \eea
%Putting all together we have poles of $C_J$ as
%\bea & & P_{K\subset J}^2,~~~~~~~~ P_{K\subset \O J}^2,~~~~~\Spab{n|(P_J+p_1)(P_{K\subset J}-P_J)(P_{K\subset J}+p_1)|1},\nn
%& & \Spab{n|P_J|1} ,~~~~~~ \Spab{n|(P_J+p_1)(P_{K\subset \O J}+P_J+p_1+p_n)(P_{K\subset \O J}+p_n)|1}~~~\label{CJ-pole}\eea
%Among these poles, some of them can be degenerated. For example, when $P_J$ is just a particle $p_i$, we have
%\bea \Spab{n|P_J|1} & \to & \Spaa{n|j}\Spbb{j|1} \eea
%where $\Spaa{n|j},\Spbb{j|1}$ are indeed undetectable by deformation $\Spab{1|n}$.

%%%%%%%%%%%%%%%%%%%%%%%%%%%%%%%%%%%%%%%%%%%

\end{document}